\documentstyle[12pt]{article}
\def\lapproxeq{\lower .7ex\hbox{$\;\stackrel{\textstyle <}{\sim}\;$}}
\def\gapproxeq{\lower .7ex\hbox{$\;\stackrel{\textstyle >}{\sim}\;$}}

\def\as{\alpha_s}
\def\asmz{\alpha_s(M_Z2)}

\def\GeV{{\rm GeV}}
\def\MeV{{\rm MeV}}
\def\qq{q \bar{q}}

\def\gl{\tilde{g}}

\def\mgl{m_{\gl}}

\def\asQ{\alpha_s(Q2)}

\def\lmsb{\Lambda_{\overline{\rm MS}}}
\def\msbar{{\overline{\rm MS}}}
\def\lmsbiv{\Lambda_{\overline{\rm MS}}{(4)}}

\begin{document}
\begin{titlepage}
\vspace*{-1cm}
\begin{flushright}
DTP/93/36 \\
RAL-93-032\\
May 1993
\end{flushright}
\vskip 1.cm
\begin{center}
{\Large\bf Light gluinos in high--$Q2$ deep inelastic scattering}

\vskip 1.cm
{\large R.G. Roberts}
\vskip .2cm
{\it
Rutherford Appleton Laboratory,  \\
Chilton, Didcot  OX11 0QX, England
} \\
\vskip .4cm
and
\vskip   .4cm
{\large  W.J. Stirling}
\vskip .2cm
{\it Department of Physics, University of Durham \\
Durham DH1 3LE, England }\\
\vskip 1cm
\end{center}

\begin{abstract}
A slight incompatibility in recent low-energy and high-energy $\as$
measurements can be  interpreted as evidence for new light colour degrees
of freedom. Assuming that these are the gluinos of a supersymmetric
extension of the Standard Model, we investigate to what extent they change the
standard QCD predictions for deep inelastic structure functions, and in
particular whether thay can be detected in such measurements at HERA.
We present a modified set of parton distributions which includes a light
gluino distribution and which can be used for further phenomenological
investigations.
\end{abstract}

\vfill
\end{titlepage}
\newpage

Much attention has been focussed on $\as$ measurements in the last few years,
motivated in part by the implications of the value of $\asmz$ for
coupling constant unification and possible hints of
supersymmetry at relatively low scales \cite{SUSY}.
A large part of the debate has centred on the issue of
which processes provide the most accurate measurements, the
reliability of  the quoted error values, and so on.
Fig.^1 shows a recent compilation  \cite{BETHKE}
of $\as$ measurements, plotted at the
`typical' energy scale $Q$ of the particular process.
{}From this compilation, an interesting
point emerges: there is a {\it hint} of a disagreement between
$\as$ values measured at low energies and those at high energies.
The solid line in Fig.^1 corresponds to $\asQ$ evaluated and evolved
at next-to-leading order in the $\msbar$ scheme,
\begin{eqnarray}
{4\pi\over\alpha_s} + {\beta_1\over\beta_0}
 \log\left({ \beta_02 \alpha_s\over
 4\pi\beta_0 + \beta_1\alpha_s}\right)
&  = & \beta_0 \log {Q2\over\Lambda2} ,  \nonumber \\
\beta_0 =  11 - {\textstyle{2\over 3}} n_f \; , \quad & &
\beta_1 =  102 - {\textstyle{38\over 3}} n_f \; ,
\label{eqa}
\end{eqnarray}
 with $\lmsbiv = 230\ \MeV$, a value consistent with all fixed-target
deep inelastic experiments and related processes \cite{MRS2}.
 We see that when extrapolated to higher $Q$ values, the coupling tends to
lie below the high-energy measurements.
Even allowing for the most
optimistic estimates of the errors \cite{BETHKE}, it is clear from Fig.^1
that there is
{\it no overall significant deviation from a unique value of} $\lmsb$.
There is only a slight hint at an incompatibility.
This notwithstanding, it has recently been argued \cite{KUHN,ENR,CLAVELLI}
 that a possible explanation of the
mismatch in the evolved and measured high-energy couplings is that at some
intermediate scale a new coloured degree of freedom is being excited, whose
effect is to slow the running of $\as$. A light supersymmetric gluino ($\gl$)
has been suggested as a possible candidate.
As the (Majorana) gluino mass threshold is crossed
the $\beta$-function coefficients change:
\begin{equation}
\beta_i \;  \rightarrow \;  \beta_i\;
 + \; \Delta\beta_i \; \theta(Q-2 m_{\gl} )\; ,
\end{equation}
where $\Delta\beta_0 = -2 $ and $ \Delta\beta_1 = -48 $.
 With $\mgl = 5\ \GeV$, the coupling
evolves as the dashed line in Fig.^1, and consistency with the high-energy
measurements is restored.
It is not our purpose here to
discuss in detail the theoretical and experimental consistency or otherwise of
this hypothesis: a  discussion can be found in \cite{ENR}. We are simply
interested to see whether such a gluino can be detected at HERA.
In our analysis, we  assume the nominal value $ \mgl = 5\ \GeV$,
although of course taken literally
the $\as$ measurements would allow a range of masses of this order.\footnote{
Note that the comparison in Fig.^1 of the running coupling including
the gluino with the LEP measurements is anyway too naive, since the presence
of the gluino  influences to some extent  the extracted $\alpha_s$ values,
particularly those from jet rates. This is discussed in detail in \cite{ENR}.}
The basic
logic is that the gluino is heavy enough to largely decouple from fixed-target
deep inelastic scattering, while at the same time light
enough to allow the coupling
to evolve to a significantly higher value at $Q \sim M_Z$  than the standard
QCD value.

The HERA high-energy $ep$ collider will, over the next few years, provide
precision measurements of the proton structure functions at scales up to
$Q2 \sim 105\ \GeV2$. Since the light gluino introduced above is
presumably electroweak neutral, its impact is expected to be very small.
This was confirmed several years ago in references
\cite{CAMP,KOUNNAS,JONES}, where
the modified evolution equations including light gluinos were presented
for leading and next-to-leading order respectively.
In fact, the present work can be regarded as an update of \cite{KOUNNAS},
taking into account (i) the increased precision of modern parton distribution
analyses, (ii) the hint from $\as$ measurements that $\mgl \sim 5\ \GeV$,
and (iii) the precise kinematic range relevant to HERA.
Thus,  if we consider the evolution of the quark, gluon  and gluino
distributions at leading order
\begin{equation}
\frac{d q(x,Q2)}{d\log Q2} = \frac{\alpha_s(Q2)}{2\pi}
\int_0x \frac{dy}{y}\left[ q(y,Q2)\;P_{qq}(\frac{x}{y})
+g(y,Q2)\;P_{qg}(\frac{x}{y}) \right]
\label{evoq}
\end{equation}
\begin{equation}
\frac{d g(x,Q2)}{d\log Q2} = \frac{\alpha_s(Q2)}{2\pi}
\int_0x \frac{dy}{y} \left[ q(y,Q2)\;P_{gq}(\frac{x}{y})
+g(y,Q2)\;P_{gg}(\frac{x}{y})+\tilde g (y,Q2)\;P_{g\tilde g}
(\frac{x}{y}) \right]
\label{evog}
\end{equation}
\begin{equation}
\frac{d \gl(x,Q2)}{d\log Q2} = \frac{\alpha_s(Q2)}{2\pi}
\int_0x \frac{dy}{y} \left[ g(y,Q2)\;P_{\gl g}(\frac{x}{y})+
\gl (y,Q2)\;P_{\gl\gl}(\frac{x}{y}) \right]
\label{evogl}
\end{equation}
with the relevant splitting functions given in reference \cite{CAMP},
we see that the only impact on the quarks (which are measured directly
by $F_2(x,Q2)$ for example) is through the coupling $\asQ$.
In fact if we assume that there  is no associated light squark, then the
leading
{\it direct} contribution to $F_2$ comes from  the $O(\as2)$   process
$\gl \gamma* \to \gl \qq$. The same process also gives a next-to-leading
contribution to the longitudinal structure function $F_L$.

To investigate the effects of the light gluino more quantitatively we have
repeated the  next-to-leading order
parton distribution analysis of reference \cite{MRS2} but now including
a light gluino with $\mgl = 5\ \GeV$. Since the bulk of the fixed-target
deep inelastic data
is below the nominal gluino threshold of $Q2 = 4 \mgl2 = 100\ \GeV2$,
there is essentially no change to the previous fits. For definiteness
we base our study on the MRS-D$_0'$ fit with $\lmsbiv = 230\ \MeV$.
Fig.^2 compares
the evolution of $F_2$ as a function of $Q2$ at fixed $x$ values
with and without a  light gluino.  The HERA kinematic limit is also shown.
In the $x \sim 0.01 - 0.1$ region where the $Q2$ evolution is weakest the
effect is very small. Only at high $x$ and high $Q2$ is there any
discernible effect, but still the maximum deviation is only of order a few
percent.
The only hope would be to compare  a precise $O(1\%)$ $F_2$ measurement
at high $x$ and $Q2$ with a
standard QCD fit evolved from lower energy
 deep inelastic data.  Even then, any uncertainty on $\lmsb$
will  effect the accuracy of the extrapolation. As an illustration,
Fig.^1 also shows (dashed lines)
the  ratio of two $F_2$'s:  one corresponding to the standard
MRS-D$_0'$  partons with $\lmsbiv = 230 \ \MeV$ and another based on
a fit with the `$+ 1 \sigma$' value $\lmsbiv = 280 \ \MeV$.
We see that at the highest $Q2$ values
the effect of changing $\lmsbiv$ by this amount is of the same order
as the effect of the light gluino, although there is a clear difference
in the shape of the evolution at lower $Q2$ values.

Unlike the quark distributions and structure functions, the evolution of
the gluon distribution {\it is} changed at leading order above threshold
 by the light gluino, Eq.^(\ref{evog}).
Unfortunately the size of the change is much smaller than the uncertainty
in the gluon from any conceivable present or future measurement. This is
illustrated in Fig.^3, which shows the standard MRS-D$_0'$ gluon evolved
to $Q2 = 5120 \ \GeV2$ with and without a $\mgl = 5\ \GeV$ gluino.
Also shown is the gluino distribution itself. Note that in
calculating this we adopt exactly  the same threshold philosophy as for
heavy quarks, {\it i.e.} we assume that the distribution is zero for
$Q2 \leq 4 \mgl2$ and evolves thereafter as if the parton was massless,
Eq.^(\ref{evogl}). This
procedure gives a reasonable description  of the structure function
data on the charm quark \cite{MRSCHARM}.  With the gluon and gluino
distributions of Fig.^3 one could, for example, investigate the changes
to  the cross sections for
such processes as large $p_T$ jet production at hadron colliders.

By momentum conservation, a non-zero gluino distribution implies a reduction
in the fraction of momentum carried by the other partons.  This
is illustrated in Fig.^4, where the momentum fraction carried by the
quarks, gluon and gluino are shown as functions of $Q2$, and for comparison,
the momentum fractions without the gluino (dashed lines).
The gluino momentum fraction increases steadily with $Q2$, reaching
$5\%$ at $Q2 = 104\ \GeV2$.

In conclusion, we find that the effect of a light gluino on the evolution
of the structure functions at HERA is minimal, being comparable to the
uncertainty of $\lmsb$ from analyses of present data. Thus attempts to
detect light gluinos at HERA should rather concentrate on the
analysis of $3+1$ jet events, with contributions from
 processes such as $\gamma q \to q\gl\gl$.
 This is analogous to searching for the process $Z\to q \bar q \gl\gl$
 in 4 jet events at LEP \cite{GLENNYS}.
Finally, we have only examined the consequences of allowing
just one light SUSY particle to modify the evolution of $F_2$. If the
gluino turns out to be really light, other SUSY particles may be light
enough to further modify the $\beta$-function at the high $Q2$ values
relevant to HERA.

\bigskip

\bigskip
\newpage
\noindent{\Large\bf Figure Captions}

\begin{itemize}
\item[{[1]}] A compilation of $\as$ measurements, taken from reference
\cite{BETHKE}. The solid line is the standard next-to-leading order QCD
coupling for $\lmsbiv = 230\ \MeV$. The dashed line corresponds to
a gluino with $\mgl = 5\ \GeV$ and the same value of $\lmsbiv$.

\item[{[2]}] Evolution in $Q2$ of the structure function $F_2$
 with a  $5\ \GeV$ gluino, compared to the standard MRS-D$_0'$ prediction
\cite{MRS2}, for various $x$ values.
Also shown (dashed lines) are the corresponding ratios
 for a structure function fitted
to low-energy data with $\lmsb = 280\ \MeV$.

\item[{[3]}] The  gluon and gluino distributions as functions of $x$
at $Q2 = 5120\ \GeV2$. The dashed line is the standard MRS-D$_0'$ gluon
with no gluino.

\item[{[4]}] Momentum fractions carried by the quarks, gluon and gluino
as functions of $Q2$. The dashed lines are the standard MRS-D$_0'$
fractions with no gluino.

\end{itemize}

\end{document}